%% file: paper.tex
\documentclass[a4paper,10pt]{article}
\pdfoutput=1
\usepackage[bookmarks=false]{hyperref}
\usepackage[numbers,sort&compress]{natbib}
\usepackage[margin=1in]{geometry}

\usepackage{amsmath}
\usepackage{amsfonts}
\usepackage{amssymb}
\usepackage{amsthm}

\usepackage[noadjust]{cite}

\usepackage{algorithmic}
\usepackage{url}
\usepackage{booktabs}
\usepackage{tikz,pgfplots}
\pgfplotsset{compat=newest}
\usetikzlibrary{external}
\tikzsetexternalprefix{tikzfigures/}
\usepgfplotslibrary{fillbetween}

\usepackage[ruled, vlined, nofillcomment]{algorithm2e}
\usepackage{algorithmic}

\SetCommentSty{mycommfont}
\SetAlgoCaptionSeparator{:}

\usepackage{framed}

\newcommand{\Transp}{\mathsf{T}}

\newcommand{\expq}{\exp_\mathsf{q}}
\newcommand{\expR}{\exp_\mathsf{R}}

\newcommand{\oriError}{\eta}

\newcommand{\ie}{i.e.\ }
\newcommand{\eg}{e.g.\ }

\usepackage{empheq}

\newcommand{\Chapterref}[1]{\hyperref[#1]{Chapter~\ref*{#1}}}
\newcommand{\Sectionref}[1]{\hyperref[#1]{Section~\ref*{#1}}}
\newcommand{\Theoremref}[1]{\hyperref[#1]{Theorem~\ref*{#1}}}
\newcommand{\Exampleref}[1]{\hyperref[#1]{Example~\ref*{#1}}}
\newcommand{\Figureref}[1]{\hyperref[#1]{Fig.~\ref*{#1}}}
\newcommand{\Tableref}[1]{\hyperref[#1]{Table~\ref*{#1}}}
\newcommand{\Algorithmref}[1]{\hyperref[#1]{Alg.~\ref*{#1}}}
\newcommand{\Stepref}[1]{\hyperref[#1]{Step~\ref*{#1}}}
\newcommand{\Assumptionref}[1]{\hyperref[#1]{Assumption~\ref*{#1}}}
\newcommand{\Definitionref}[1]{\hyperref[#1]{Definition~\ref*{#1}}}
\newcommand{\Appendixref}[1]{\hyperref[#1]{Appendix~\ref*{#1}}}

\providecommand{\keywords}[1]{\textbf{\textit{Keywords: }} #1}

\usepackage{color}

\newlength\figureheight 
\newlength\figurewidth

\newcommand{\mytilde}{\raise.17ex\hbox{$\scriptstyle\mathtt{\sim}$}}

\graphicspath{{./figures/}}

\begin{document}
\input{coverArXiv.tex}

\title{A Fast and Robust Algorithm \\ for Orientation Estimation using Inertial Sensors}

\author{Manon~Kok$^\star$ and Thomas B. Sch\"on$^{\star \star}$ \\
\small{$^{\star}$Delft Center for Systems and Control, Delft University of Technology, the Netherlands, e-mail: m.kok-1@tudelft.nl} \\
\small{$^{\star \star}$Department of Information Technology, Uppsala University, Sweden, e-mail: thomas.schon@it.uu.se} \\
}
\date{\empty}

\section{Introduction}
Orientation estimation using inertial and magnetometer measurements is by now a well-studied problem with applications in \eg human motion analysis~\citep{kokHS:2014,miezalTB:2016} and robotics~\citep{corkeLD:2007,atchuthan:2018}. In recent years, there has been an increasing demand for algorithms that are computationally inexpensive and that can estimate orientation in real-time on a microprocessor, \eg illustrated by the wide-spread use of the techniques from~\citep{madgwick:2010,madgwickHV:2011}. These algorithms open up for real-time human motion analysis~\citep{angKR:2004,karatsidisRKNSBHV:2018} and control of robots~\citep{atchuthan:2018,kanazawaNKKKOIY:2015}. In this work we present a novel computationally efficient algorithm for orientation estimation. Our algorithm's robustness against inaccuracies in the accelerometer and magnetometer measurement models adds to its practical usefulness.

There exists a vast range of orientation estimation algorithms. Their differences lie both in the estimation method and in the parametrisation of the orientation. For example, both~\citep{madgwick:2010,madgwickHV:2011} and the extended Kalman filter (EKF) presented in~\citep{sabatini:2006} estimate the orientation parametrised as a unit quaternion. Normalisation of these unit quaternions is necessary for the estimates to remain valid orientations. This normalisation introduces errors especially when large updates to the estimates are made, for instance due to low sampling rates or large initialisation errors~\citep{julierV:2007,kokHS:2017}. Alternatively, algorithms can estimate orientation deviation from a linearisation point, see \eg \citep{roetenbergSV:2007,roetenberg:2006}. An example is provided by the multiplicative EKF (MEKF) that parametrises the orientation deviation in terms of a rotation vector (axis-angle)~\citep{kokHS:2017,markley:2003,crassidisMC:2007}. A similar approach is often used in robotics~\citep{bloeschEtAl:2016,barfoot:2016,grisettiKSB:2010,grisettiKSFH:2010,forsterCDS:2016}. This formulation avoids the issues with quaternion normalisation while at the same time reducing the state dimension.

In our algorithm, we make similar design choices as the widely used filter published by Madgwick et al.~\citep{madgwick:2010,madgwickHV:2011}. Because of this, our filter inherits its desirable properties such as easy tuning and accurate estimates also in the presence of model errors. However, since we estimate the orientation in terms of an orientation deviation parametrised using a rotation vector, our filter reduces the computational complexity by approximately 1/3 and obtains more accurate estimates when large updates to the estimates are made. 

\section{Background on sensor models}
\label{sec:models}
Our algorithm makes use of standard sensor models for orientation estimation, hence ensuring the wide applicability of our method. First of all, we model the gyroscope measurements $y_{\omega,t}$ as 
\begin{align}
y_{\omega,t} &= \omega_t + e_{\omega,t},
\label{eq:gyrMeasModel}
\end{align}
where $\omega_t$ denotes the angular velocity and $e_{\omega,t}$ the measurement noise. Note that the sensor model~\eqref{eq:gyrMeasModel} can optionally be extended with a gyroscope bias. 

We assume that the sensor's acceleration $a^\text{n}$ is approximately zero and that the accelerometer therefore only measures the Earth's gravity $g^\text{n}$. The superscript $n$ explicitly indicates that the vector is expressed in the navigation frame $n$, which is aligned with Earth’s gravity and the local (Earth's) magnetic field. The accelerometer measurements $y_{\text{a},t}$ are hence modelled as 
\begin{align}
y_{\text{a},t} = R(q^{\text{bn}}_t) \left( a^\text{n} - g^\text{n} \right) + e_{\text{a},t} \approx - R(q^{\text{bn}}_t)  g^\text{n} + e_{\text{a},t}. \label{eq:accMeasModel} 
\end{align}
We represent the orientation at time $t$ using a unit quaternion denoted by $q^{\text{nb}}_t$. The superscript $nb$ indicates that the quaternion represents the rotation from the body frame $b$, which is aligned with the sensor axes, to the navigation frame $n$. The operation $R(q^{\text{nb}}_t)$ converts the quaternion to a rotation matrix. The reverse orientation used in~\eqref{eq:accMeasModel} is given by $R(q^{\text{nb}}_t) = ( R(q^{\text{bn}}_t) )^\Transp$. Due to the fact that the Earth's gravity depends on the sensor's location and since the unit in which the sensor expresses the acceleration varies, we assume that $y_{\text{a},t}$ has unit norm and define $g^\text{n} = \begin{pmatrix} 0 & 0 & 1 \end{pmatrix}^\Transp$. To this end, we preprocess the accelerometer measurements before using them in the measurement model~\eqref{eq:accMeasModel}. 

We assume that the magnetometer measures a local, constant (Earth's) magnetic field $m^\text{n}$. The magnetometer measurements $y_{\text{m},t}$ are therefore modelled as
\begin{align}
y_{\text{m},t} &= R(q^\text{bn}_t)  m^\text{n} + e_{\text{m},t}. \label{eq:magMeasModel}
\end{align} 
Because the unit in which the sensor expresses the magnetic field measurement varies, we normalise the magnetometer measurements before using them in the model~\eqref{eq:magMeasModel}. Furthermore, we assume that $m^\text{n} = \begin{pmatrix} \cos \delta & 0 & -\sin \delta \end{pmatrix}^\Transp$, where $\delta$ is the local dip angle. Alternatively, there also exist methods to estimate the local magnetic field from data, see \eg \citep{madgwick:2010}. 

Note that the noise terms $e_{\text{a},t}$ and $e_{\text{m},t}$ consist of both the sensor noise as well as model errors. These model errors are for instance due to non-zero acceleration or due to the presence of ferromagnetic material in the vicinity of the sensor. 

\section{Fast and Robust Orientation Estimation}
\label{sec:oriEst}
\begin{figure}
\includegraphics[width = 1\columnwidth,trim={0 12.8cm 16.5cm 0},clip]{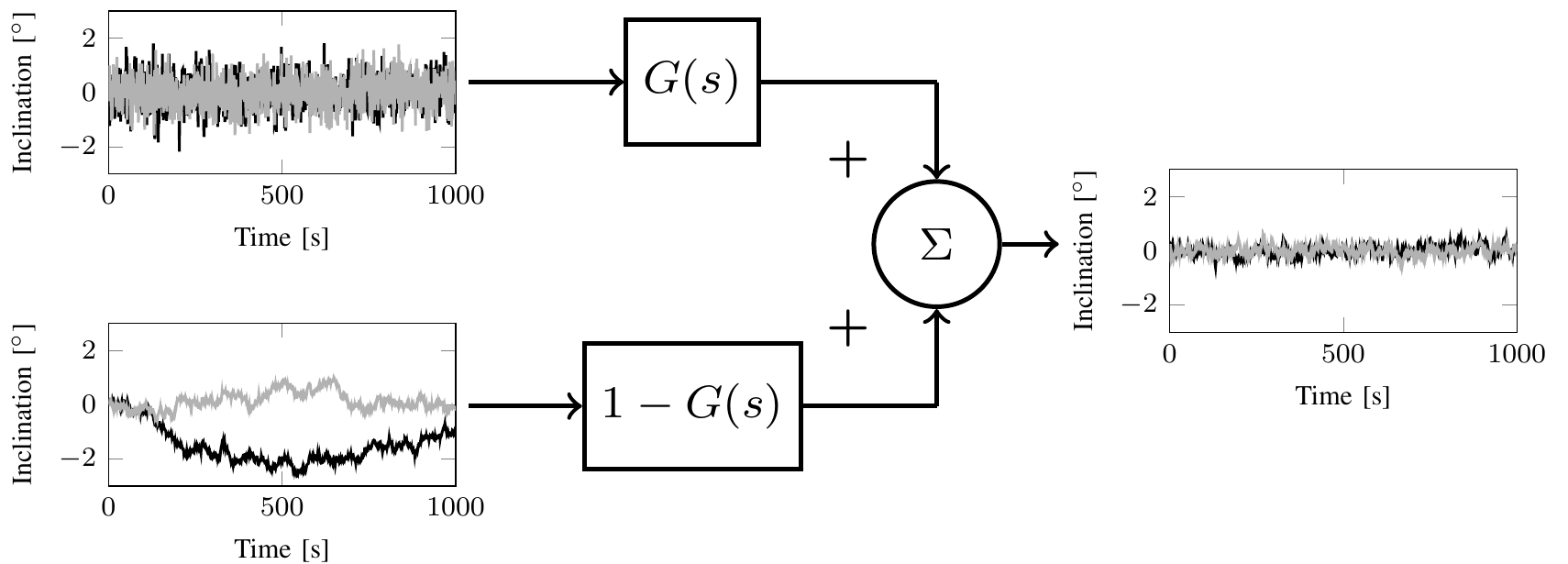}
\caption{Illustration of the complementary filter for inclination estimation using simulated data. The inclination (roll in black, pitch in grey) from the accelerometer (top left) is low-pass filtered while the inclination from the gyroscope (bottom left) is high-pass filtered to obtain the resulting inclination estimate (right).}
\label{fig:cf}
\end{figure}
Assuming well-calibrated sensors and a known initial orientation, it is possible to estimate the sensor orientation based only on the gyroscope measurements. This estimate, which we denote by $q^\text{nb}_\omega$, is accurate on a short time scale but drifts over longer time horizons. On the other hand, the orientation can be estimated using the accelerometer and magnetometer measurements. This estimate, denoted by $q^\text{nb}_\text{am}$, is less accurate than $q^\text{nb}_\omega$ on a short time scale but does not drift. These complementary properties can be exploited using a \emph{complementary filter} in which $q^\text{nb}_\text{am}$ is low-pass filtered while $q^\text{nb}_\omega$ is high-pass filtered~\citep{brown:1972,higgins:1975}. This can be written as 
\begin{align}
\hat{Q}^\text{nb}(s) = G(s) Q^\text{nb}_\text{am}(s) + (1-G(s)) Q^\text{nb}_\omega(s),
\label{eq:cf-laplace}
\end{align}
where $s$ denotes the Laplace variable. Furthermore, the transfer function $G(s)$ is given by $G(s) = \tfrac{1}{a s + 1}$, $Q^\text{nb}(s)$ denotes the orientation $q^\text{nb}$ in the Laplace domain and $\hat{q}^\text{nb}$ is the resulting filtered orientation. This process is visualised in Figure~\ref{fig:cf}. 

Discretising~\eqref{eq:cf-laplace} using backward Euler gives 
\begin{align}
\hat{q}^\text{nb}_t = (1-\gamma_t) q^\text{nb}_{\text{am},t} + \gamma_t \left(\hat{q}^\text{nb}_{t-1} + T \omega_{\text{q},t} \right), 
\label{eq:cf-discrete}
\end{align}
where $T$ denotes the sampling time, $\gamma_t = \tfrac{a}{a+T}$ and $\omega_{\text{q},t}$ represents the angular velocity expressed in terms of a quaternion. At first glance, \eqref{eq:cf-discrete} might cause concern because $\hat{q}^\text{nb}_t$ will not be a valid rotation since this quaternion is no longer normalised. It will, however, become clear in the remainder of this section that the deviation from the unit norm will be small due to the high sampling rates of the sensors. This deviation can be resolved by normalising $\hat{q}^\text{nb}_t$. Although this leads to minor inaccuracies, it is fairly common practice in many orientation estimation algorithms, see \citep{kokHS:2017} and references therein.

\subsection{Orientation from gyroscope measurements}
The angular velocity measured by the gyroscope can be used to model the dynamics of the orientation as \citep{kokHS:2017,gustafsson:2012}
\begin{align}
q^\text{nb}_{t} &= q^\text{nb}_{t-1} \odot \expq \left( \tfrac{T}{2} y_{\omega,t} \right) \approx q^\text{nb}_{t-1} + \tfrac{T}{2} S(q_{t-1}^\text{nb}) y_{\omega,t},
\label{eq:dynModel}
\end{align}
where $\odot$ denotes the quaternion product and $\expq$ denotes the quaternion version of the vector defined as 
\begin{align}
\expq(y) &= \begin{pmatrix} \cos{\alpha} & v^\Transp \sin{\alpha} \end{pmatrix}^\Transp, \quad \alpha = \| y \|_2, \quad v = \tfrac{y}{\alpha}. 
\end{align}
Furthermore, for $q = \begin{pmatrix} q_0 & q_1 & q_2 & q_3 \end{pmatrix}^\Transp = \begin{pmatrix} q_0 & q_v^\Transp \end{pmatrix}^\Transp$,
\begin{align}
S(q) = \begin{pmatrix} -q_v \\ q_0 \, \mathcal{I}_3 - [q_v \times] \end{pmatrix},
\label{eq:Sq}
\end{align}
where $[\, \cdot \, \times]$ denotes the matrix cross product and $\mathcal{I}_3$ denotes the identity matrix of size 3. Comparing~\eqref{eq:cf-discrete} and~\eqref{eq:dynModel}, we have that the angular velocity expressed in terms of a quaternion is given by
\begin{align}
\omega_{\text{q},t} = \tfrac{1}{2} S(\hat{q}_{t-1}^\text{nb}) y_{\omega,t}.
\label{eq:doriGyr}
\end{align}

\subsection{Orientation from accelerometer and magnetometer}
Estimating the orientation from accelerometer and magnetometer measurements is a widely known problem, see \eg~\citep{wahba:1965,markleyM:2000}. It can be formulated  as an optimisation problem 
\begin{align}
\min_{\oriError_t} V(\oriError_t) = 
 \min_{\oriError_t} \tfrac{1}{2} \| y_{\text{a},t} + \left( \expR(\oriError_t) \right)^\Transp R(\tilde{q}^\text{bn}_t) g^\text{n} \|_2^2 + 
\tfrac{1}{2} \| y_{\text{m},t} - \left( \expR(\oriError_t) \right)^\Transp R(\tilde{q}^\text{bn}_t) m^\text{n} \|_2^2,
\label{eq:optAccMag}
\end{align}
where $\| \cdot \|_2$ denotes the two-norm. In~\eqref{eq:optAccMag} we use the measurement models~\eqref{eq:accMeasModel} and~\eqref{eq:magMeasModel} but write the orientation in terms of a linearisation point and an associated deviation as 
\begin{subequations}
\begin{align}
R(q_t^\text{nb}) &= R(\tilde{q}_t^\text{nb}) \expR(\oriError_t), \\
\expR(\oriError_t) &= \mathcal{I}_3 + \sin{\alpha} \left[ v \times \right] + \left( 1 - \cos{\alpha} \right) \left[ v \times \right]^2 \nonumber \\
&\approx \mathcal{I}_3 + [\oriError_t \times], \label{eq:expRapprox}
\end{align}
\end{subequations}
where the approximation in~\eqref{eq:expRapprox} assumes small $\oriError_t$. Rewriting the problem in this way allows us to optimise over an orientation deviation parametrised in terms of a rotation vector~\citep{kokHS:2017,shuster:1993}, rather than optimising over a unit quaternion. We therefore avoid issues with quaternion normalisation. Furthermore, the number of optimisation variables reduces from four to three. Our approach draws inspiration from the multiplicative extended Kalman filter (MEKF)~\citep{kokHS:2017,markley:2003,crassidisMC:2007} and from approaches within the field of robotics \citep{bloeschEtAl:2016,barfoot:2016,grisettiKSB:2010,grisettiKSFH:2010,forsterCDS:2016}.

Inspired by~\citep{madgwick:2010,madgwickHV:2011}, instead of solving~\eqref{eq:optAccMag} for each time step, we perform only a \emph{single} gradient descent iteration. This results in a significant computational speed-up and because of the high sampling rates of the sensors, the corrections that need to be made are typically minor and the estimates will converge over time. Linearising $V(\eta_t)$ from~\eqref{eq:optAccMag} around $\tilde{q}_{t}^\text{bn} = \hat{q}_{t-1}^\text{bn}, \oriError_{t} = 0$ using~\eqref{eq:expRapprox}, the gradient descent step is given by
\begin{subequations}
\begin{align}
\hat{\oriError}_t &= - \mu_t \nabla V(\oriError_t), \\ 
\nabla V(\oriError_t) &=  - [R(\hat{q}_{t-1}^\text{bn}) g^\text{n} \times] \left( y_{\text{a},t} + R (\hat{q}_{t-1}^\text{bn}) g^\text{n} \right) + 
[R(\hat{q}_{t-1}^\text{bn}) m^\text{n} \times] \left( y_{\text{m},t} - R (\hat{q}_{t-1}^\text{bn}) m^\text{n} \right), \label{eq:gradientAccMag}
\end{align}%
\label{eq:gradientDescentAccMag}%
\end{subequations}%
where $\mu_t$ is the gradient descent step length. The estimate $\oriError_t$ can subsequently be used to compute $q^\text{nb}_{\text{am},t}$ from~\eqref{eq:cf-discrete} as 
\begin{align}
q_{\text{am},t}^\text{nb} &= \hat{q}_{t-1}^\text{nb} \odot \expq \left( \tfrac{1}{2} \hat{\oriError}_t \right) \approx \hat{q}_{t-1}^\text{nb} + \tfrac{1}{2} S(\hat{q}_{t-1}^\text{nb}) \hat{\oriError}_t.
\label{eq:oriAccMag}%
\end{align}

\subsection{Resulting algorithm}
Inserting~\eqref{eq:oriAccMag} and~\eqref{eq:doriGyr} into~\eqref{eq:cf-discrete}, we obtain
\begin{align}
\hat{q}_t^\text{nb} = \hat{q}_{t-1}^\text{nb} + \tfrac{1}{2} S(\hat{q}_{t-1}^\text{nb}) \left( \gamma_t T y_{\omega,t} - \mu_t (1-\gamma_t) \nabla V(\oriError_t) \right).
\label{eq:oriCF-2}
\end{align}
It now remains to choose $\gamma_t$ and $\mu_t$. Similarly to~\citep{madgwick:2010}, we choose $\gamma_t \approx 1$. In other words, we mainly rely on the integration of the gyroscope measurements, but use the accelerometer and magnetometer measurements to correct for the integration drift illustrated in Fig.~\ref{fig:cf}. This choice is motivated by the fact that the orientation estimates obtained from the accelerometer and the magnetometer are typically more noisy than those obtained using the gyroscope measurements (see Fig.~\ref{fig:cf}). Furthermore, the accelerometer and the magnetometer measurement models~\eqref{eq:accMeasModel} and~\eqref{eq:magMeasModel} are often violated due to acceleration of the sensor or the presence of magnetic disturbances.

Similarly to~\citep{madgwick:2010}, we choose the scaling factor of the gradient descent direction, $\mu_t (1-\gamma_t)$, equal to $\tfrac{\beta T}{\| \nabla V(\oriError_t) \|}$. We will in Section~\ref{sec:properties} illustrate that scaling the step with the norm $\| \nabla V(\oriError_t) \|$ results in an algorithm that is quite robust against violations of the accelerometer and magnetometer measurement models. The choice of $\beta$ depends on the amount of drift expected from integration of the gyroscope noise. In the case of Gaussian noise on the gyroscope data, $e_{\omega,t} \sim \mathcal{N}(0,\sigma_\omega^2)$, integration of the gyroscope measurements in one dimension results in an integration drift distributed as $T e_{\omega,t} \sim \mathcal{N}(0,T^2 \sigma_\omega^2)$. Using the fact that the gyroscope measurements are a three-dimensional vector and are integrated according to~\eqref{eq:dynModel}, the standard deviation of the integration drift on the unit quaternion is given by $\sqrt{3} \sigma_\omega T$. This is therefore a reasonable choice for $\beta T$ resulting in a good compromise. The resulting filter equations can now be written as 
\begin{subequations}
\begin{align}
\hat{q}^\text{nb}_{t} &\approx \hat{q}^\text{nb}_{t-1} + \tfrac{T}{2} S(\hat{q}^\text{nb}_{t-1}) \, \hat{\omega}_t, \\
\hat{\omega}_t &= y_{\omega,t} - \beta  \tfrac{\nabla V(\oriError_t)}{\| \nabla V(\oriError_t) \|}, 
\end{align}
\label{eq:oriEst}%
\end{subequations}%
and the resulting solution is summarised in Alg.~\ref{alg:oriEst}. As can be seen in~\eqref{eq:oriEst}, we directly estimate the angular velocity which is subsequently used to update the orientation. This has close connections to the approaches discussed in~\citep{sjanicSG:2017,skoglundSK:2017}. 

\section{Numerical Illustrations}
\label{sec:properties}
In this section we numerically illustrate the properties of Alg.~\ref{alg:oriEst} and compare them to the filter from~\citep{madgwick:2010,madgwickHV:2011} and to an MEKF implemented as described in~\citep{kokHS:2017}. To this end, we run 100 Monte Carlo simulations, each consisting of 8000 samples during which the sensor is first stationary for 200 samples and then consecutively rotates 360 degrees around each axis in 200 samples per rotation axis. This movement is repeated 10 times. The sampling time is set to 10 Hz. For ease of interpretation, we decouple the magnetometer and accelerometer information by assuming that the measurements are collected on the equator, \ie the dip angle $\delta$ is zero and hence $m^\text{n} = \begin{pmatrix} 1 & 0 & 0 \end{pmatrix}^\Transp$.

\begin{algorithm}[t]
  \caption{Fast and Robust Orientation Estimation}
  \label{alg:oriEst}
  \begin{minipage}{\linewidth-14.45pt}
  \begin{algorithmic}[1]
    \REQUIRE Gyroscope measurements $y_{\omega,t}$, normalised accelerometer and magnetometer measurements $y_{\text{a},t}$ and $y_{\text{m},t}$, sampling time $T$, tuning parameter $\beta$ and the orientation estimate at the previous time instance $\hat{q}^\text{nb}_{t-1}$. 
    \ENSURE Orientation estimate $\hat{q}^\text{nb}_t$.
    \STATE Compute $\nabla V(\oriError_t)$ from~\eqref{eq:gradientAccMag} using $y_{\text{a},t}$, $y_{\text{m},t}$ and $\hat{q}^\text{nb}_{t-1}$.
    \STATE Obtain the updated orientation estimate $\hat{q}^\text{nb}_{t}$ from~\eqref{eq:oriEst} using $\nabla V(\oriError_t)$, $\beta$, $y_{\omega,t}$, $T$, $\hat{q}^\text{nb}_{t-1}$ and $S(q)$ from~\eqref{eq:Sq}.
  \end{algorithmic}
  \end{minipage}
\end{algorithm}

\subsection{Computational complexity}
One of the widely known benefits of the filter from~\citep{madgwick:2010,madgwickHV:2011} is its low computational complexity. More specifically, the filter uses only 218 arithmetic operations per filter iteration. Instead of directly estimating the orientation parametrised as a unit quaternion, we estimate the angular velocity as described in~\eqref{eq:oriEst}, effectively reducing the state dimension from~4 to~3. This reduces the number of arithmetic operations per filter iteration to 140, a reduction of $36\%$. In Table~\ref{tab:results} we show the average computational time per filter iteration in our simulations for a Matlab implementation run on a 3.1 GHz Intel Core i5 processor. As can be seen, Alg.~\ref{alg:oriEst} is indeed $36\%$ faster than the filter from~\citep{madgwick:2010,madgwickHV:2011}. Note that our MEKF implementation has not been optimised for computational speed but is known to be slower than both other filters. 

\label{subsec:compCompl}
\subsection{Gaussian noise with known characteristics}
\label{subsec:gaussian}
We first consider an idealised case where the measurement noises are Gaussian with known covariances and the initial sensor orientation is known. More specifically, we set the standard deviation of the gyroscope noise to $\sigma_\omega = \tfrac{5\pi}{180}$ rad/s, and that of the normalised accelerometer and magnetometer noise to $\sigma_\text{a} = \sigma_\text{m} = 0.01$. For Alg.~\ref{alg:oriEst} we choose $\beta$ as explained in Section~\ref{sec:oriEst}. We tune the filter from~\citep{madgwick:2010,madgwickHV:2011} similarly (setting the tuning parameter in that filter to $\sqrt{\tfrac{3}{4}} \sigma_\omega$). Note that different values of these parameters did not improve the performance of the algorithms. The MEKF is expected to outperform the other two algorithms since the update equation of an EKF is typically closer to optimal than the update equation based on a normalised gradient descent step. The latter update strategy is used in both Alg.~\ref{alg:oriEst} and the filter from~\citep{madgwick:2010,madgwickHV:2011}. As can be seen from Table~\ref{tab:results}, the MEKF outperforms the other two algorithms only by a small amount. Alg.~\ref{alg:oriEst} and the filter from~\citep{madgwick:2010,madgwickHV:2011} perform more or less equally.

\subsection{Accelerometer and magnetometer model inaccuracies}
Inaccuracies in the accelerometer and magnetometer models occur regularly in practice, since the acceleration of the sensor is seldom exactly zero, as assumed in the measurement model~\eqref{eq:accMeasModel}, and since the magnetic field is often disturbed due to the presence of ferromagnetic material. To analyse the sensitivity of the three algorithms to these model inaccuracies, we consider the same scenario as in Section~\ref{subsec:gaussian}, but randomly replace $5\%$ of the normalised accelerometer and magnetometer data with outliers. These outliers are sampled from a Gaussian distribution with covariance equal to the identity matrix. As can be seen in Table~\ref{tab:results}, Alg.~\ref{alg:oriEst} and the filter from~\citep{madgwick:2010,madgwickHV:2011} are more robust than the MEKF and barely suffer from the outliers in the data. This is caused by the normalised gradient descent update step. Note that the robustness of the MEKF can be improved by using outlier rejection or by using techniques from \eg \citep{kassamP:1985,gandhiM:2010}, which is outside the scope of this work. 

\subsection{Large orientation uncertainties}
\begin{figure}
\centering
\caption{Orientation errors for the first 150 samples of 100 Monte Carlo simulations with a fixed initial orientation error. The mean and spread (2 std) are shown for the MEKF (blue), Alg.~\ref{alg:oriEst} (black) and the filter from~\citep{madgwick:2010,madgwickHV:2011} (red).}
\label{fig:oriError}
\includegraphics[scale=1]{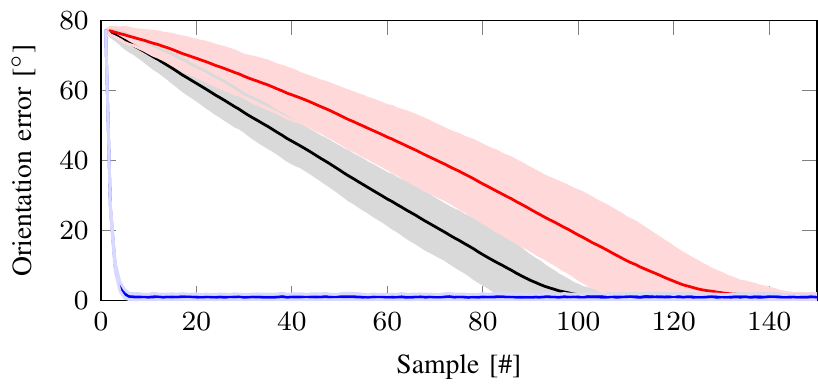}
\end{figure}

In practice, orientation estimates are occasionally very uncertain. This can be due to large initialisation errors or due to sensors not providing measurements for an extended period of time. When accurate orientation information subsequently becomes available, filtering algorithms need some time to recover from these large orientation uncertainties. The amount of time this takes depends both on the update strategy of the algorithm as well as on linearisation errors. To study the behaviour of the three algorithms for this case, we again consider the scenario from Section~\ref{subsec:gaussian} but assume that there is a large uncertainty in the initial orientation. We visualise the orientation errors of the three algorithms for the first 150 samples for 100 Monte Carlo simulations with a fixed initial orientation error in Fig.~\ref{fig:oriError}. As can be seen, the MEKF consistently recovers much faster from an erroneous initialisation due to its adaptive update strategy. However, Alg.~\ref{alg:oriEst} converges faster than the filter from~\citep{madgwick:2010,madgwickHV:2011} after an erroneous initialisation. This difference can be attributed to the fact that estimation of the angular velocity as in~\eqref{eq:oriEst} avoids linearisation errors that occur when directly estimating the orientation parametrised as a unit quaternion.

In conclusion, Alg.~\ref{alg:oriEst} inherits the desirable robustness against accelerometer and magnetometer outliers of the filter from~\citep{madgwick:2010,madgwickHV:2011} as illustrated in Table~\ref{tab:results}. Furthermore, it reduces the computational complexity with $36\%$ and converges faster after large orientation errors as illustrated in Fig.~\ref{fig:oriError}.

\begin{table}[!t]
\caption{RMSE and computational times from the numerical analysis.}
\label{tab:results}
\centering
\begin{tabular}{c c c l l l}
\toprule
 & & Roll & Pitch & Yaw & Time/iter \\
 \midrule      
Known noise & Alg.~\ref{alg:oriEst} & $0.71^\circ$ & $0.66^\circ$ & $0.71^\circ$ & $\textbf{6.40}~\mu \text{s}$  \\ 
variances and & \citep{madgwick:2010,madgwickHV:2011} & $0.72^\circ$ & $0.65^\circ$ & $0.71^\circ$ & $10~\mu \text{s}$\\ 
initial orientation & MEKF & $\textbf{0.66}^\circ$ & $\textbf{0.60}^\circ$ & $\textbf{0.66}^\circ$ & $83.9~\mu \text{s}$ \\ 
 \midrule    
 5\% outliers & Alg.~\ref{alg:oriEst} & $\textbf{0.77}^\circ$ & $\textbf{0.72}^\circ$ & $\textbf{0.77}^\circ$ & $\textbf{6.40}~\mu \text{s}$\\ 
magnitude  & \citep{madgwick:2010,madgwickHV:2011} & $0.78^\circ$ & $\textbf{0.72}^\circ$ & $0.78^\circ$ & $10~\mu \text{s}$\\ 
$\mathcal{N}(0,\mathcal{I})$ & MEKF & $9.03^\circ$ & $6.37^\circ$ & $9.10^\circ$ & $83.9~\mu \text{s}$ \\  
\midrule
Experimental  & Alg.~\ref{alg:oriEst} & $\textbf{0.69}^\circ$ & $\textbf{0.43}^\circ$ & $\textbf{0.36}^\circ$ & $\textbf{6.40}~\mu \text{s}$ \\ 
data & \citep{madgwick:2010,madgwickHV:2011} & $\textbf{0.69}^\circ$ & $0.44^\circ$ & $\textbf{0.36}^\circ$ & $10~\mu \text{s}$ \\ 
 & MEKF & $0.79^\circ$ & $0.46^\circ$ & $0.41^\circ$ & $83.9~\mu \text{s}$ \\ 
\bottomrule 
\end{tabular}
\end{table}

\section{Experimental Results}
To validate Alg.~\ref{alg:oriEst} on experimental data, we use 30 seconds of inertial and magnetometer data collected at 100 Hz using a Trivisio Colibri Wireless IMU~\citep{trivisio}. The data is collected in a lab equipped with multiple cameras~\citep{vicon} that are able to track optical markers to obtain highly accurate ground truth reference orientation information, against which we can compare our estimates. We time-synchronise and align the data as described in~\citep{kokHS:2017}. The gyroscope bias has been estimated based on a stationary portion of data and the data has been corrected for this. Furthermore, the initial orientation is estimated based on the first accelerometer and magnetometer samples~\citep{kokHS:2017}. The RMSE of the estimates of the three algorithms with respect to the reference data can be found in Table~\ref{tab:results}. These results were obtained using $\beta = 2.4 \cdot 10^{-3}$ for Alg.~\ref{alg:oriEst}, by setting the tuning parameter from the filter from~\citep{madgwick:2010,madgwickHV:2011} to $1.4 \cdot 10^{-3}$, and by setting the process, accelerometer and magnetometer noise covariances in the MEKF to $1.3 \cdot 10^{-3} \, \mathcal{I}_3$, $2.63 \cdot 10^{-2} \, \mathcal{I}_3$ and $2.5 \cdot 10^{-2} \, \mathcal{I}_3$, respectively. These were experimentally found to be good values.

\section{Conclusion}
We have presented a novel algorithm for online, real-time orientation estimation. The algorithm reduces the computational complexity by $36\%$ compared to the approach from~\citep{madgwick:2010,madgwickHV:2011}, which is widely known for its low computational complexity. It is more robust against outliers than an extended Kalman filter implementation and reduces the issues related to quaternion normalisation compared to~\citep{madgwick:2010,madgwickHV:2011}, resulting in better convergence in the case of large orientation errors. Our new algorithm has also been shown to obtain good results on experimental data. The source code is available on \url{github.com/manonkok/fastRobustOriEst}.

\section*{Acknowledgments}
This research was partially supported by the Swedish Foundation for Strategic Research (SSF) via the project \emph{ASSEMBLE} (contract number: RIT15-0012) and by the project \emph{Learning flexible models for nonlinear dynamics} (contract number: 2017-03807), funded by the Swedish Research Council. High accuracy reference measurements are provided through the use of the Vicon real-time tracking system courtesy of the UAS Technologies Lab, Artificial Intelligence and Integrated Computer Systems Division (AIICS) at the Department of Computer and Information Science (IDA), Link\"oping University, Sweden \url{http://www.ida.liu.se/divisions/aiics/aiicssite/index.en.shtml}. The authors would like to thank Fredrik Olsson for comments and suggestions that greatly improved this paper. 

\bibliographystyle{unsrtnat}
\bibliography{references}

\end{document}

%% file: coverArXiv.tex

\newcommand{\coverTitle}{A Fast and Robust Algorithm \\ for Orientation Estimation \\using Inertial Sensors}
\newcommand{\coverYear}{2019}

\newcommand{\coverAuthors}{Manon~Kok$^{\star}$ and Thomas~B.~Sch\"on$^{\star \star}$ \\ \vspace{3mm}
\small{$^\star$Delft Center for Systems and Control, Delft University of Technology, the Netherlands, e-mail: m.kok-1@tudelft.nl} \\
\small{$^{\star \star}$Department of Information Technology, Uppsala University, Sweden, e-mail: thomas.schon@it.uu.se} 
}

\begin{titlepage}
\begin{center}
%

\vspace*{2.5cm}
%
{\Huge \bfseries \coverTitle  \\[0.4cm]}

%
{\Large \coverAuthors \\[1.5cm]}

\renewcommand\labelitemi{\color{red}\large$\bullet$}
\begin{itemize}
\item {\Large \textbf{Please cite this version:}} \\[0.4cm]
\normalsize
Manon Kok and Thomas B. Sch\"on, ``A Fast and Robust Algorithm for Orientation Estimation using Inertial Sensors", IEEE Signal Processing Letters, 2019. DOI: 10.1109/LSP.2019.2943995
\end{itemize}

\end{center}

\vspace{1cm}

\begin{abstract}
\noindent We present a novel algorithm for online, real-time orientation estimation. Our algorithm integrates gyroscope data and corrects the resulting orientation estimate for integration drift using accelerometer and magnetometer data. This correction is computed, at each time instance, using a single gradient descent step with fixed step length. This fixed step length results in robustness against model errors, \eg caused by large accelerations or by short-term magnetic field disturbances, which we numerically illustrate using Monte Carlo simulations. Our algorithm estimates a three-dimensional update to the orientation rather than the entire orientation itself. This reduces the computational complexity by approximately 1/3 with respect to the state of the art. It also improves the quality of the resulting estimates, specifically when the orientation corrections are large. We illustrate the efficacy of the algorithm using experimental data. 
\end{abstract}

\keywords{Orientation estimation, inertial sensors, complementary filter, multiplicative extended Kalman filter.}

\vfill

\end{titlepage}